\documentclass[prc,showpacs,amsmath, showkeys, twocolumn]{revtex4}
\usepackage{graphicx}
\usepackage{txfonts}
\usepackage{bm}
\usepackage{graphicx,amsmath,mathrsfs}
\usepackage{amssymb}
\usepackage[section]{placeins}
\usepackage{multirow}
\usepackage{CJK}
\usepackage{tabularx}
\usepackage{array}
\usepackage{color}
\usepackage{epstopdf}

\newcommand{\PreserveBackslash}[1]{\let\temp=\\#1\let\\=\temp}
\newcolumntype{C}[1]{>{\PreserveBackslash\centering}p{#1}}
\newcolumntype{R}[1]{>{\PreserveBackslash\raggedleft}p{#1}}
\newcolumntype{L}[1]{>{\PreserveBackslash\raggedright}p{#1}}

\usepackage{ulem}
\newcommand{\red}[1]{\textcolor[rgb]{1.00,0.00,0.00}{#1}}

\newcommand{\delete}{\bgroup\markoverwith{\textcolor{red}{\rule[0.5ex]{2pt}{1pt}}}\ULon}

\renewcommand{\emph}[1]{\red{#1}}

\newcommand{\lrlc}[1]{\left|#1\right>}
\newcommand{\lrcl}[1]{\left<#1\right|}
\newcommand{\lrb}[1]{\left(#1\right)}
\newcommand{\lrs}[1]{\left[#1\right]}

\newcommand{\svec}[1]{{\mbox{\boldmath${ #1}$}}}

\newcommand{\err}[1]{{\footnotesize#1}}

\begin{document}

\title{Descriptions of Carbon isotopes within relativistic Hartree-Fock-Bogoliubov theory}
\author{Xiao Li Lu }
\affiliation{School of Nuclear Science and Technology, Lanzhou University, 730000 Lanzhou, China}
\author{Bao Yuan Sun }
\affiliation{School of Nuclear Science and Technology, Lanzhou University, 730000 Lanzhou, China}
\author{Wen Hui Long }\email{longwh@lzu.edu.cn}
\affiliation{School of Nuclear Science and Technology, Lanzhou University, 730000 Lanzhou, China}

\begin{abstract}
Within the relativistic Hartree-Fock-Bogoliubov (RHFB) theory, the structure properties of Carbon isotopes are systematically studied. To provide better overall description, the finite-range Gogny force D1S with an adjusted strength factor is adopted as the effective paring interaction in particle-particle channel. The self-consistent RHFB calculations with density-dependent meson-nucleon couplings indicate the single-neutron halo structures in both $^{17}$C and $^{19}$C, whereas the two-neutron halo in $^{22}$C is not well supported. It is also found that close to the neutron drip line there exists distinct odd-even staggering on neutron radii, which is tightly related with the blocking effects and correspondingly the blocking effect plays a significant role in the single-neutron halo formation.                                                                                                                                                                                                                                                                                 \end{abstract}

\pacs{21.60.Jz, 21.10.Dr, 21.30.Fe, 21.60.-n}
\keywords{Carbon isotopes, Relativistic Hartree-Fock-Bogoliubov, Single-neutron halo, Blocking effects}

\maketitle 

\section{Introduction}
During the past decades, the radioactive ion beams (RIBs) have greatly extended our knowledge of nuclear physics, from which are obtained the critical data for nuclear physics, astrophysics,  as well as for testing the standard model. With worldwide and rapid development of RIB facilities, the investigations of the nuclear systems under extreme conditions generate new frontiers in nuclear physics. For example, the exotic nuclei \cite{Chulkov1996, Mueller2001, Tanihata1995, John1999} have fascinated more and more interests due to the unexpected exotic modes therein. One of the representatives is the nuclear halo structure characterized by a dilute matter distribution contributed by several (in general two) loosely bound valence neutrons (or protons) surrounding a condensed core, which was first found in $^{11}$Li \cite{Tanihata1985}. As the typical light nuclei, the Carbon isotopes have been devoted many efforts to probing the possible halo structure \cite{Bazin1998, Fang2004, Horiuchi2006} and specifically recent measured reaction cross section of $^{22}$C \cite{Tanaka2010} seems to assert a new two-neutron halo structure, which has also attracted fairly large interests from the community \cite{Sharma2011, LuLu2012, Fortune2012}.

In fact, the exotic modes keeping found in the weakly bound nuclear systems also bring serious challenges on the reliability of the nuclear theoretical models. When extending to the limit of stability of isotopes or isotones, the single neutron or proton separation energies become comparable to the pairing gap energy, such that the continuum effects can be easily involved by pairing correlations and play a significant role in determining the structure properties of exotic nuclei \cite{Meng1996, Meng1998, Meng1998NPA}. In terms of Bogoliubov quasi-particle, the relativistic Hartree-Bogoliubov (RHB) theory \cite{Meng1998NPA, Vretenar2005, Meng2006} has unified the descriptions of relativistic Hartree (RH) mean field and pairing correlations, and consequently the continuum effects are involved automatically. Since the first self-consistent description of nuclear halo structure in $^{11}$Li \cite{Meng1996}, the RHB theory has been successfully applied in predicting the giant halos in Ca \cite{Meng2002, Long2010} and Zr \cite{Meng1998, Grasso2006, Long2010} isotopes, as well as the restoration of relativistic symmetry \cite{Meng1999PRC} and superheavy magic structures \cite{Zhang2005}.

With the inclusion of Fock terms in the mean field, the relativistic Hartree-Fock-Bogoliubov (RHFB) theory with density-dependent meson-nucleon couplings \cite{Long2010RHFB} provides a new self-consistent platform for the exploration of exotic nuclei, e.g., predicting the giant halos in Cerium isotopes \cite{Long2010}. In addition, the inclusion of Fock terms has brought substantial improvements in the self-consistent description of nuclear shell structures \cite{Long2007} and the evolutions \cite{Long2008, Long2009},  the relativistic symmetry restorations \cite{Long2006, Long2007, Liang2010}, and the low-energy excitation modes \cite{Liang2008}.

In this work, the structure properties of Carbon isotopes, particularly the possible halo structures therein, will be studied systematically within the RHFB and RHB theories. The contents are organized as follows. In the Sec. \ref{Theory}, we introduce the general formalism of the RHFB equations with finite range (Gogny) pairing force. In Sec. \ref{result} the discussions are concentrated on the halo structures and odd-even staggering (OES) on the neutron radii for Carbon isotopes. Finally, a brief summary and perspective are given in Sec. \ref{summary}.

\section{Theoretical framework and numerical details}\label{Theory}
In relativistic nuclear models the effective force between the nucleons is mediated by the exchange of mesons and photons. Based on that, the model Lagrangian contains the system degrees of freedom associated with the nucleon $\psi$, isoscalar scalar $\sigma$-meson, isoscalar vector $\omega$-meson, isovector vector $\rho$-meson, isovector pseudo-scalar $\pi$-meson and the photon ($A$) fields \cite{Bouyssy1987, Long2007}. Following the standard variational procedure, one can get the equations of motion for nucleons, mesons, and photons, namely the Dirac, Klein-Gordon, and Proca equations, as well as the continuity equation for energy-momentum tensor, from which is derived the system Hamiltonian. In the terms of the creation and annihilation operators $(c_\alpha^\dag,c_\alpha)$ defined by the stationary solutions of the Dirac equation, the Hamiltonian operator can be expressed as
\begin{align}
H=&\sum_{\alpha\beta}c_\alpha^\dag c_\beta T_{\alpha\beta}+\frac{1}{2}\sum_{\alpha\alpha'\beta\beta'}c_\alpha^\dag c_\beta^\dag c_{\beta'}c_{\alpha'}\sum_{\phi}V_{\alpha\beta\alpha'\beta'}^\phi,\label{H}
\end{align}
where $T_{\alpha\beta}$ is the kinetic energy and the two-body terms $V_{\alpha\beta\alpha'\beta'}^\phi$correspond with the meson- (or photon-) nucleon couplings denoted by $\phi$,
\begin{align}
T_{\alpha\beta}&=\int d\bm{r}\bar{\psi}_\alpha(\bm{r})(-i\svec\gamma\cdot\svec\nabla+M)\psi_\beta(\bm{r}),\\
V_{\alpha\beta\alpha'\beta'}^\phi&=\int d\bm{r}d\bm{r}'\bar{\psi}_\alpha(\bm{r})\bar{\psi}_\beta(\bm{r}')\Gamma_\phi(\svec r,\svec r')\nonumber\\
&\hspace{8em}\times D_\phi(\bm{r},\bm{r}')\psi_{\beta'}(\bm{r}')\psi_{\alpha'}(\bm{r}).
\end{align}
In above equations, $\Gamma_\phi(x,x')$ represent the interaction matrices associated with the $\sigma$-scalar, $\omega$-vector, $\rho$-vector, $\rho$-tensor, $\rho$-vector-tensor, $\pi$-pseudo-vector and photon-vector couplings, and $D_\phi(\bm{r}, \bm{r}')$ denotes relevant meson (photon) propagator, and $M$ is the nucleon mass (for details see Refs. \cite{Bouyssy1987, Long2007, Long2010RHFB}).

Standing on the level of relativistic Hartree-Fock (RHF) approach, the contributions from the negative energy states in the Hamiltonian (1) are neglected as usual, i.e., the so-called no-sea approximation \cite{Bouyssy1987}. The Hartree-Fock ground state $\lrlc{\Phi_0} $ is then determined and consequently is derived the energy functional $E$, i.e., the expectation of Hamiltonian with respect to $\lrlc{\Phi_0} $,
\begin{align}
\lrlc{\Phi_0} =& \prod_{i=1}^A c_i^\dag |0\rangle,& E=&\lrcl{\Phi_0} H\lrlc{\Phi_0} ,
\end{align}
where the index $i$ denotes the positive energy states and $|0\rangle$ is the vacuum state. In the energy functional $E$, the two-body interactions $V^\phi$ lead to two types of contributions, i.e., the direct (Hartree) and exchange (Fock) terms. Within RHFB \cite{Long2010RHFB}, the mean field part contains both types of the contributions, i.e., the RHF approach \cite{Long2006DDRHF}, whereas within RHB the Fock terms are neglected just for simplicity.

For the open-shell nuclei, the pairing correlations, which lead to valence particles spreading over the orbits around the Fermi level, have to be taken into account. Different from simple BCS method \cite{Bardeen1957}, the Bogoliubov theory can unify the descriptions of mean field and pairing correlations in terms of Bogoliubov quasi-particle. It is of special significance in exploring the nuclei far from the $\beta$-stability line where the continuum effects become essential and the simple BCS method may break down. In the prior studies with both RHB and RHFB theories, it is already demonstrated that the scattering of the Cooper pairs into the continuum plays an essential role in the formation of the halo structures \cite{Long2010, Meng1996, Meng1998}.

Following the standard procedure of the Bogoliubov transformation \cite{Gor1958, Kucharek1991}, the RHFB equation can be derived as,
\begin{equation}
\begin{split}
\int d\svec r'& \lrb{\begin{matrix}h(\svec r, \svec r') & \Delta(\svec r, \svec r')\\[0.5em] -\Delta(\svec r, \svec r')& h(\svec r, \svec r')\end{matrix}} \lrb{\begin{matrix}\psi_U(\svec r')\\[0.5em] \psi_V(\svec r')\end{matrix}}\\&\hspace{6em} =  \lrb{\begin{matrix} \lambda + E_q &0\\[0.5em] 0&\lambda-E_q
\end{matrix}}\lrb{\begin{matrix}\psi_U(\svec r)\\[0.5em] \psi_V(\svec r)\end{matrix}},\label{RHFB}
\end{split}
\end{equation}
where $\psi_U$ and $\psi_V$ are the quasi-particle spinors, $E_q$ denotes the single quasi-particle energy, and the chemical potential $\lambda$ is introduced to keep the particle number on the average. For the single-particle Hamiltonian $h(\bm{r}, \bm{r}')$, it consists of three parts, i.e., the kinetic energy $h^{kin}$, local potential $h^D$ and non-local one $h^E$,
\begin{subequations}
\begin{align}
  h^{kin}(\bm{r}, \bm{r}') &= \gamma^0\lrb{\svec\gamma\cdot\svec p+ M}\delta(\bm{r}-\bm{r}'),\\
  h^{D}(\bm{r}, \bm{r}') &= \lrs{\Sigma_T(\bm{r})\gamma_5+\Sigma_0(\bm{r})+\gamma^0\Sigma_S(\bm{r})}\delta(\bm{r}-\bm{r}'),\\
  h^{E}(\bm{r}, \bm{r}') &= \lrb{\begin{matrix} Y_G(\svec r, \svec r') & Y_F(\svec r, \svec r') \\[0.5em] X_G(\svec r, \svec r') & X_F(\svec r, \svec r')  \end{matrix}}.
\end{align}
\end{subequations}
Detials are referred to Refs. \cite{Long2010RHFB, Bouyssy1987}. The pairing potential in the RHFB equation (\ref{RHFB}) reads as
\begin{align}
\Delta_\alpha(\bm{r},\bm{r}')=-\frac{1}{2}\sum_\beta V_{\alpha\beta}^{pp}(\bm{r},\bm{r}')\kappa_\beta(\bm{r},\bm{r}'),
\end{align}
with the pairing tensor $\kappa$
\begin{align}
\kappa_\alpha(\bm{r},\bm{r}')=\psi_{V_\alpha}(\bm{r})^*\psi_{U_\alpha}(\bm{r}').
\end{align}
For the pairing interaction $V^{pp}$, it is generally taken as a phenomenological form with great success in RHB theory \cite{Vretenar2005, Gonzalez1996} and conventional HFB theory \cite{Decharg1980, Dobaczewski1984}. In this work, we utilize the finite-range Gogny force D1S \cite{Berger1984} with additional strength factor $f$ as the effective pairing interaction,
\begin{align}
V(\bm{r},\bm{r}')=&f\sum_{i=1,2}e^{((r-r')/\mu_i)^2}(W_i+B_iP^\sigma-H_iP^\tau-M_iP\sigma P^\tau),
\end{align}
where $\mu_i,W_i,B_i,H_i,M_i(i=1,2)$ are the Gogny parameters and the factor $f$ will be adjusted to provide better overall description for the selected Carbon isotopes.

Due to the numerical difficulties originating from both RHF mean field and finite-range pairing interactions, the integro-differential RHFB equation (\ref{RHFB}) is solved by expanding the quasi-particle spinors on the Dirac Woods-Saxon (DWS) basis \cite{Shanggui2003}, which can provide appropriate asymptotic  behaviors for the continuum states in the weakly bound nuclei. For the calculations of Carbon isotopes, the DWS basis parameters are taken as follows: the spherical box-size is fixed to 30 fm and consistently the numbers of basis states with positive and negative energies are chosen as 48 and 12, respectively.

\begin{table}[h]
\caption{Details for the effective interactions PKA1, PKO2 PKO3, PKDD, DDME2, PK1 and NL2. The abbreviations DD and NL denote the density-depdent meson-nucleon couplings and the non-linear self-couplings, respectively.}\label{tab:coupling} \tabcolsep=1em
\begin{tabular}{cccccc}\hline\hline
       &  DD  & NL  & Fock term & $\pi$ & $\rho$-tensor\\ \hline
PKA1   & yes  & no  & yes       & yes   & yes          \\
PKO2   & yes  & no  & yes       & no    & no           \\
PKO3   & yes  & no  & yes       & yes   & no           \\  \hline \hline
PKDD   & yes  & no  & no        & no    & no           \\
DD-ME2 & yes  & no  & no        & no    & no           \\  \hline
PK1    & no   & yes & no        & no    & no           \\
NL2    & no   & yes & no        & no    & no           \\
\hline\hline
\end{tabular}
\end{table}

\begin{table}[h]
\caption{Blocked quasi-neutron ($\nu$) orbits of the ground states of the odd Carbon isotopes $^{15, 17, 19, 21}$C determined by the calculations of PKA1, PKO2, PKO3, PKDD, DD-ME2, PK1 and NL2.} \label{tab:bl}\tabcolsep=0.5em
\begin{tabular}{cccccccc} \hline\hline
        &PKA1          &PKO2          &PKO3          &PKDD          &DD-ME2        &PK1           &NL2          \\ \hline
$^{15}$C&$\nu s_{1/2}$ &$\nu d_{5/2}$ &$\nu d_{5/2}$ &$\nu s_{1/2}$ &$\nu s_{1/2}$ &$\nu s_{1/2}$ &$\nu s_{1/2}$\\
$^{17}$C&$\nu s_{1/2}$ &$\nu s_{1/2}$ &$\nu s_{1/2}$ &$\nu s_{1/2}$ &$\nu s_{1/2}$ &$\nu s_{1/2}$ &$\nu s_{1/2}$\\
$^{19}$C&$\nu d_{5/2}$ &$\nu s_{1/2}$ &$\nu s_{1/2}$ &$\nu s_{1/2}$ &$\nu s_{1/2}$ &$\nu s_{1/2}$ &$\nu s_{1/2}$\\
$^{21}$C&$\nu s_{1/2}$ &$\nu s_{1/2}$ &$\nu s_{1/2}$ &$\nu s_{1/2}$ &$\nu s_{1/2}$ &$\nu s_{1/2}$ &$\nu s_{1/2}$\\ \hline\hline
\end{tabular}
\end{table}

In this work, we performed systematical calculations for the Carbon isotopes from $^{10}$C to $^{22}$C by the RHFB and RHB theories, utilizing the effective interactions with density-dependent meson couplings, namely PKA1 \cite{Long2007}, PKO2 \cite{Long2008} and PKO3 \cite{Long2008}, PKDD \cite{Long2004} and DD-ME2 \cite{Lalazissis2005}, and the ones with non-linear self-couplings, i.e., PK1 \cite{Long2004} and NL2 \cite {Lee1986}. The details of the selected effective Lagrangians are referred to Table \ref{tab:coupling}. For the odd Carbon isotopes, the blocking effects are taken into account. In general, e.g., under the BCS scheme, several orbits around the Fermi surface are blocked separately and the blocking with the strongest binding corresponds to the ground state \cite{Junqing2002}. In present study, the self-consistent calculations are carried out within the Bogoliubov scheme and naturally the blocking effects are considered under the same scheme to keep the consistence of the theory itself. According to the mapping relation between the HF single-particle and Bogoliubov quasi-particle states (see Fig. 11 in Ref. \cite{Meng2006}), the blocked quasi-particle orbit can be determined as the lowest ones, e.g., the orbits $1s_{1/2}$ or $1d_{5/2}$ for $^{15,17,19,21}$C. In table \ref{tab:bl} are shown the blocking configurations for the ground states of the odd Carbon isotopes close to the neutron drip line.

\section{Results and discussion}\label{result}

\begin{table*}[htbp]
\caption{The root mean square deviations (MeV) from the data \cite{Audi2011} for the binding energies $E_b$, single- ($S_n$) and two-neutron ($S_{2n}$) separation energies of the Carbon isotopes.
The results are extracted from the calculations by PKA1, PKO2 PKO3, PKDD, DD-ME2, PK1 and NL2 with different pairing strength factor $f$. See the text for details.}\label{tab:deviation}
\tabcolsep=3pt 
\renewcommand\arraystretch{1.0}
\begin{tabular}{c|c|c|c|c|c|c|c|c|c|c|c|c|c|c|c|c|c|c|r|c|c}
  \hline\hline
\multirow{2}{*}{$f$}  & \multicolumn{3}{c|}{PKA1} & \multicolumn{3}{c|}{PKO2} & \multicolumn{3}{c|}{PKO3} & \multicolumn{3}{c|}{PKDD} & \multicolumn{3}{c|}{DD-ME2} & \multicolumn{3}{c|}{PK1} & \multicolumn{3}{c}{NL2}\\  \cline{2-22}
     & $E_b$ & $S_n$ & $S_{2n}$ & $E_b$ & $S_n$ & $S_{2n}$ & $E_b$ & $S_n$ & $S_{2n}$ & $E_b$ & $S_n$ & $S_{2n}$ & $E_b$ & $S_n$ & $S_{2n}$ & $E_b$ & $S_n$ & $S_{2n}$ & $E_b$~~ & $S_n$ & $S_{2n}$\\
 \hline
1.00 &	1.34 &	0.88 &	0.81 &	2.14 &	0.43 &	1.02 &	2.10 &	0.65 &	1.66 &	2.83 &	0.78 &	1.72 &	3.00 &	0.97 &  1.92 & 1.38 &	0.71 &	1.42 &	 12.29 &	1.77 &	3.87\\
\hline
1.10 &	1.58 &	0.78 &	0.75 &	1.78 &	0.41 &	0.84 &	1.35 &	0.59 &	1.72 &	2.46 &	0.65 &	1.59 &	2.49 &	0.78 &  1.71 & 0.90 &	0.66 &	1.28 &	 11.18 &	1.64 &	3.59\\
\hline
1.15 &	1.71 &	0.95 &	1.10 &	1.43 &	0.46 &	0.80 &	0.90 &	0.62 &	1.79 &	2.17 &	0.63 &	1.52 &	2.29 &	0.63 &  1.68 & 0.60 &	0.78 &	1.27 &	 10.54 &	1.57 &	3.43\\
\hline
1.20 &	2.08 &	1.03 &	1.22 &	1.02 &	0.55 &	0.81 &	0.51 &	0.69 &	1.80 &	1.81 &	0.66 &	1.48 &	1.90 &	0.64 &  1.73 & 0.46 &	0.80 &	1.27 &	 9.83 &	1.51 &	3.25 \\
\hline
1.25 &	2.60 &	1.14 &	1.34 &	0.65 &	0.67 &	0.99 &	0.67 &	0.79 &	1.79 &	1.38 &	0.72 &	1.57 &	1.45 &	0.68 &  1.80 & 0.81 &	0.92 &	1.44 &	 9.05 &	1.46 &	3.05 \\
\hline\hline
\end{tabular}
\end{table*}

\begin{figure*}[htbp]\ifpdf
\includegraphics[width=0.45\textwidth]{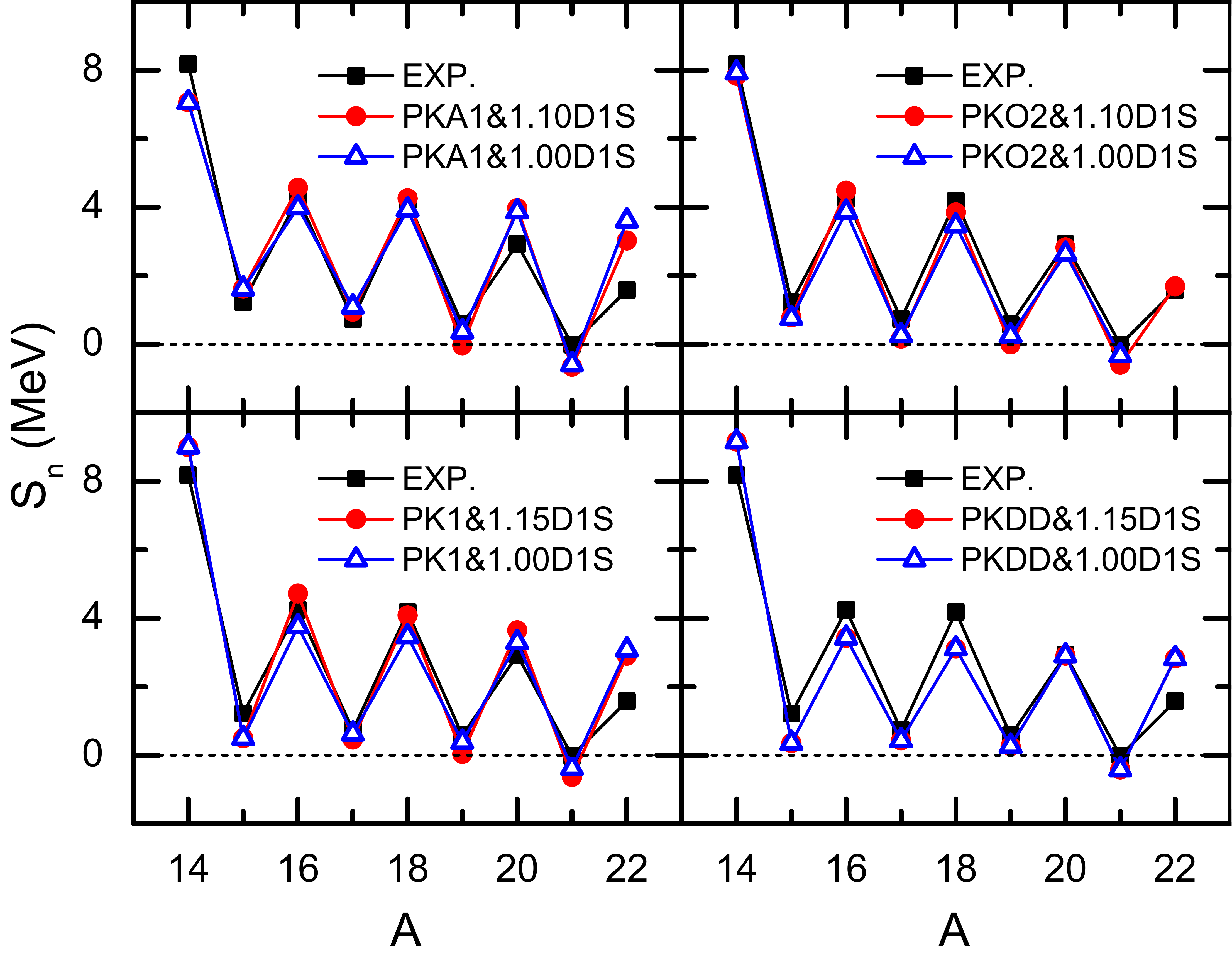}~~
\includegraphics[width=0.45\textwidth]{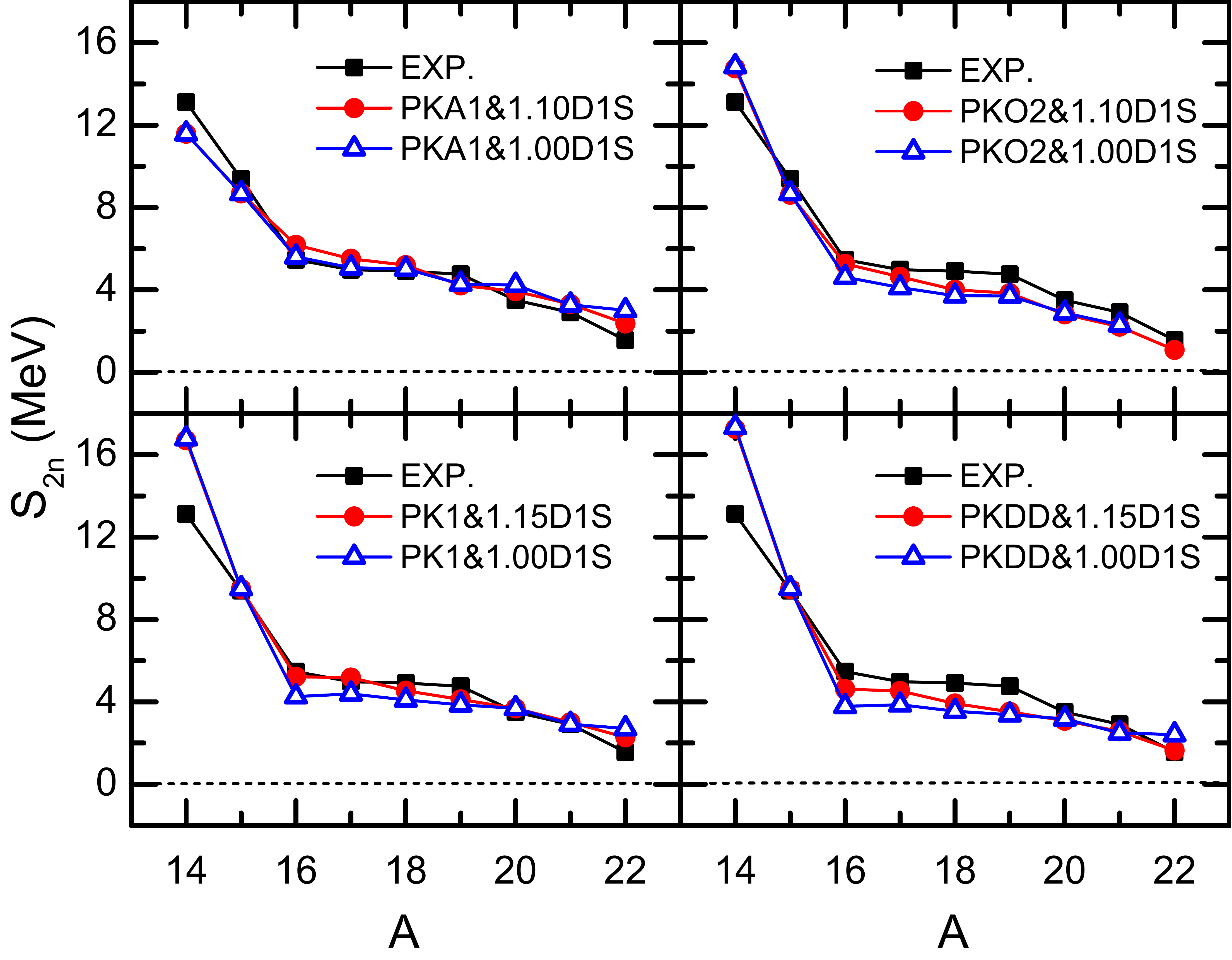}
\else
\includegraphics[width=0.45\textwidth]{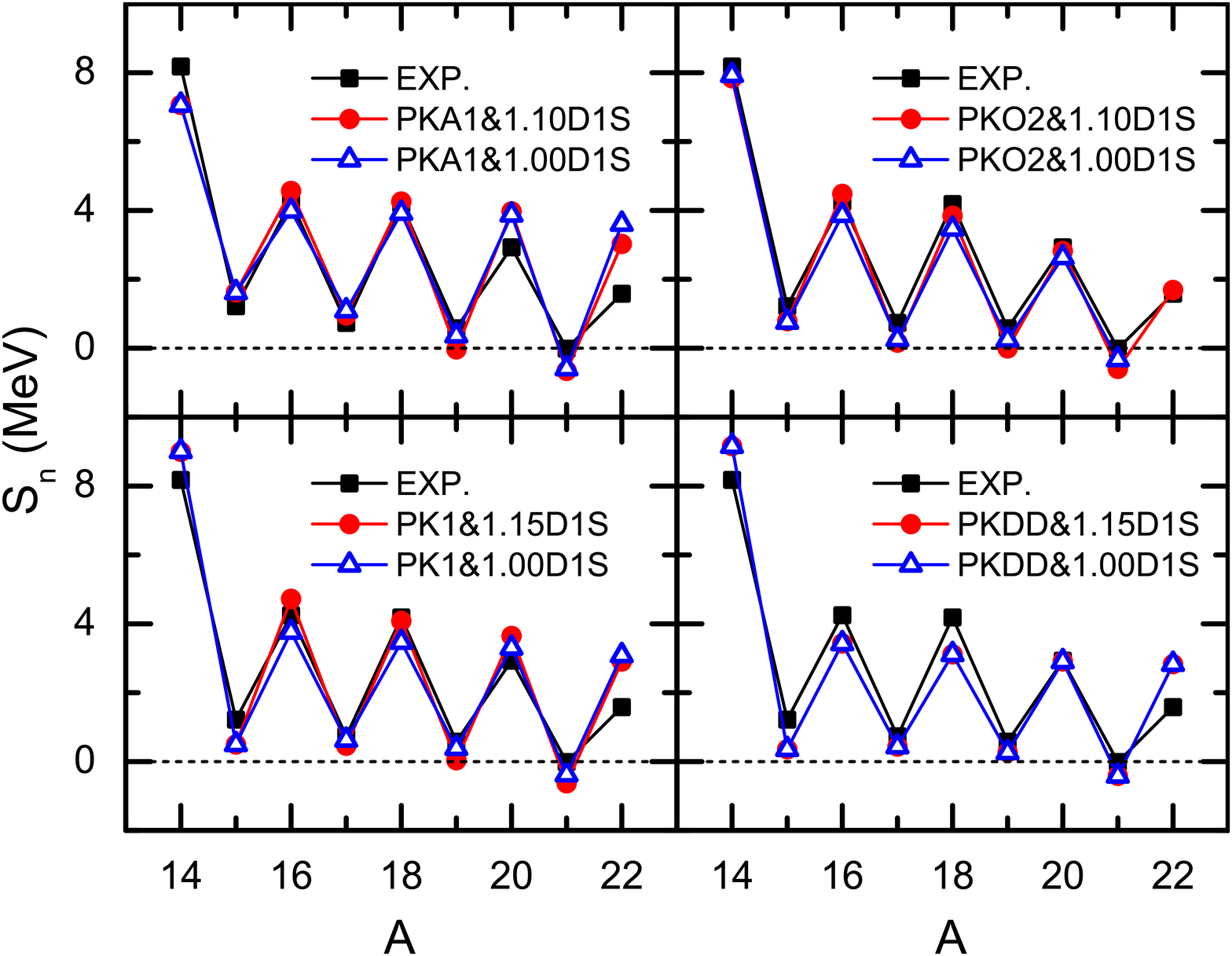}~~
\includegraphics[width=0.45\textwidth]{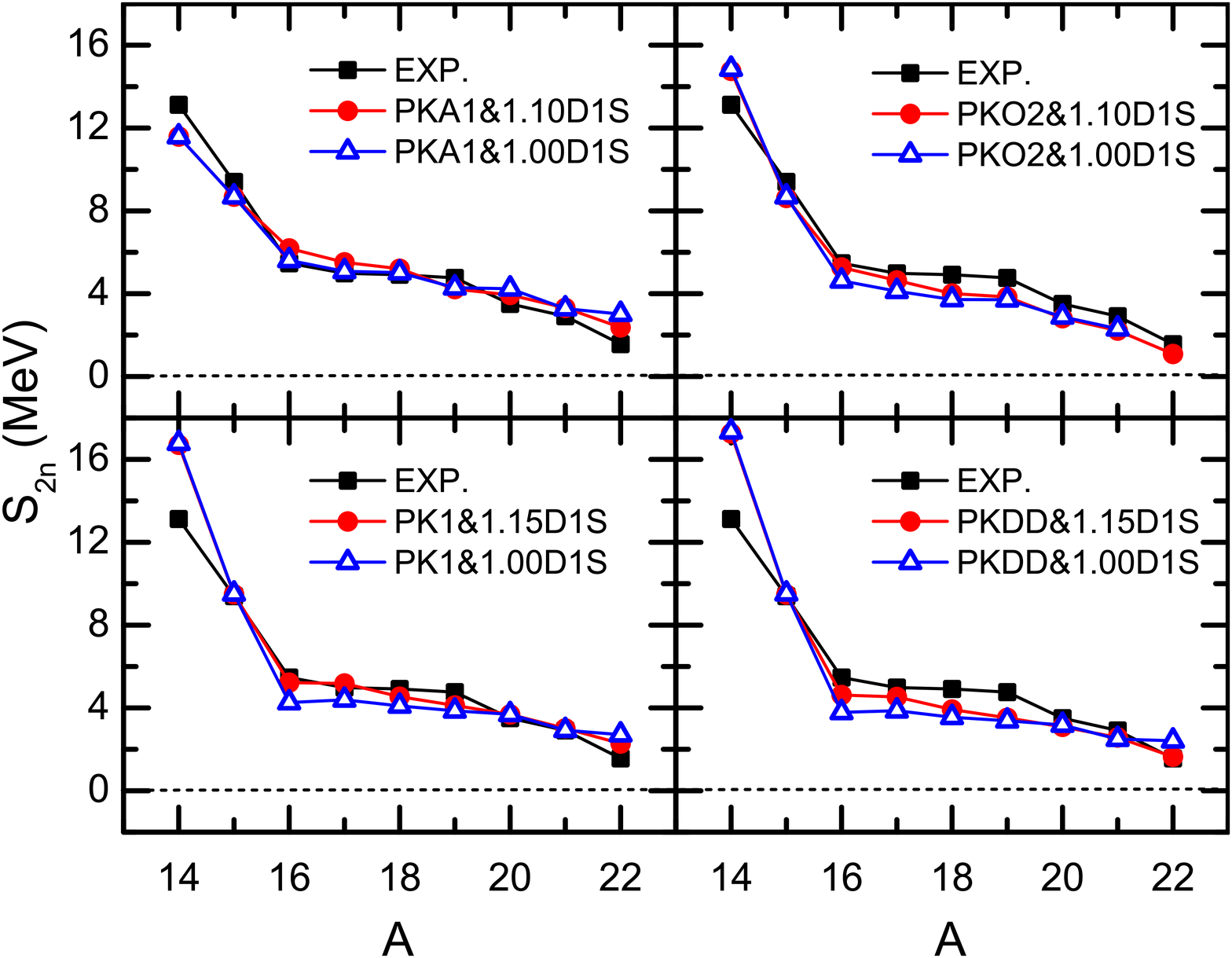}
\fi
\caption{(Color online) Single- ($S_n$: left panels) and two-neutron ($S_{2n}$: right panels) separation energies for Carbon isotopes from $^{14}$C to $^{22}$C. The results are calculated by RHFB with PKA1 and PKO2 (upper panels), and by RHB with PK1 and PKDD (lower panels). The filled circles and open up-triangles denote the results calculated by taking the Gogny force D1S with/without optimized strength factor as the effective pairing interactions, respectively. As the references, the data extracted from Ref. \cite{Audi2011} are shown in filled squares. }\label{fig:S}
\end{figure*}

To get appropriate pairing effects, firstly the systematical calculations with different pairing strength factors are performed for the Carbon isotopes. In Table \ref{tab:deviation} are shown the root mean square deviations from the data \cite{Audi2011} for the binding energies $E_b$, single- ($S_n$) and two-neutron ($S_{2n}$) separation energies extracted from the calculations with the selected effective Lagrangians in Table \ref{tab:coupling}. It is found that all the effective interactions present appropriate agreement with the data, except NL2 which fails to provide enough binding for Carbon isotopes with about 10\% deviations. It can be also seen that the systematics on single- and two-neutron separation energies are improved quantitatively with the modified pairing interactions. Referred to the single-neutron separation energy $S_n$, the optimized strength factors are determined as $f=1.1$ for PKA1, PKO2 and PKO3, and $f=1.15$ for PKDD, DD-ME2 and PK1, and $f=1.25$ for NL2. Among the selected effective Lagrangians, one can find in Table \ref{tab:deviation} that PKO2 provides the best agreement with the data on both $S_n$ and $S_{2n}$, which may imply the most reliable systematics.

Figure \ref{fig:S} presents the single-neutron separation energies $S_n$ (left panels) and two-neutron ones $S_{2n}$ (right panels) of Carbon isotopes calculated with PKA1, PKO2, PK1 and PKDD, as compared to the data (in filled squares) \cite{Audi2011}. The comparison is performed between the calculations with the original effective pairing interaction Gogny-D1S (open up-triangles) and the ones with the optimized strength factor $f$ (filled circles). From Fig. \ref{fig:S} one can find that the modification on the pairing force brings some systematical improvements on both $S_n$ and $S_{2n}$, especially for the calculations with PKA1 and PKO2. The results calculated by PKO3 and DD-ME2 are omitted because of the similar systematics as PKO2 and PKDD, respectively. Specifically with the original Gogny-D1S, PKO2 can not reproduce the drip line $^{22}$C which becomes bound with enhanced pairing force (see Fig. \ref{fig:S}). Combined with the results in Table \ref{tab:deviation}, we utilize PKO2 with optimized pairing force as the representative to analyze the detailed structure properties of Carbon isotopes in the following discussions.

\begin{figure}[htbp]
\ifpdf
\includegraphics[width=0.45\textwidth]{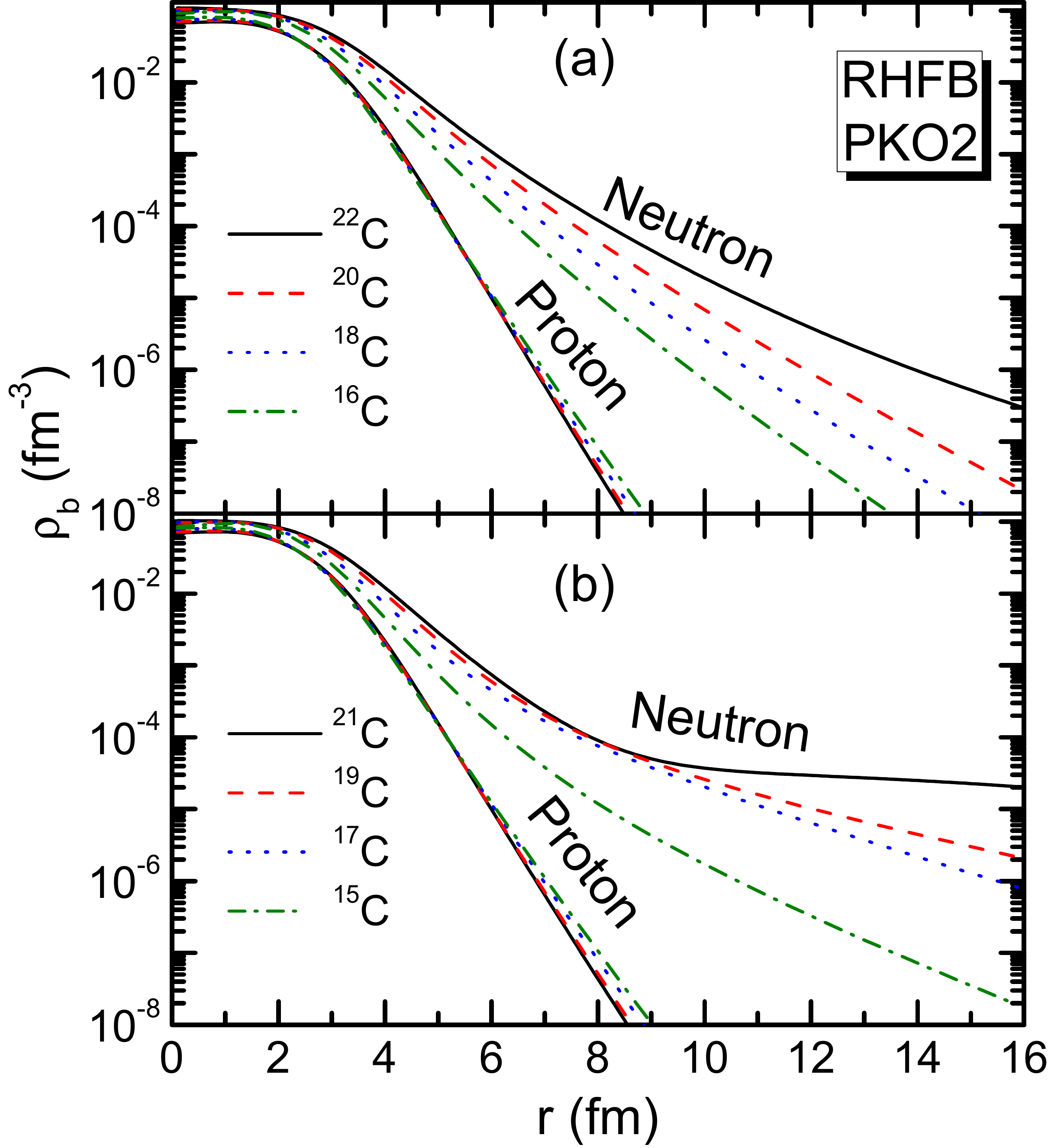}
\else
\includegraphics[width=0.45\textwidth]{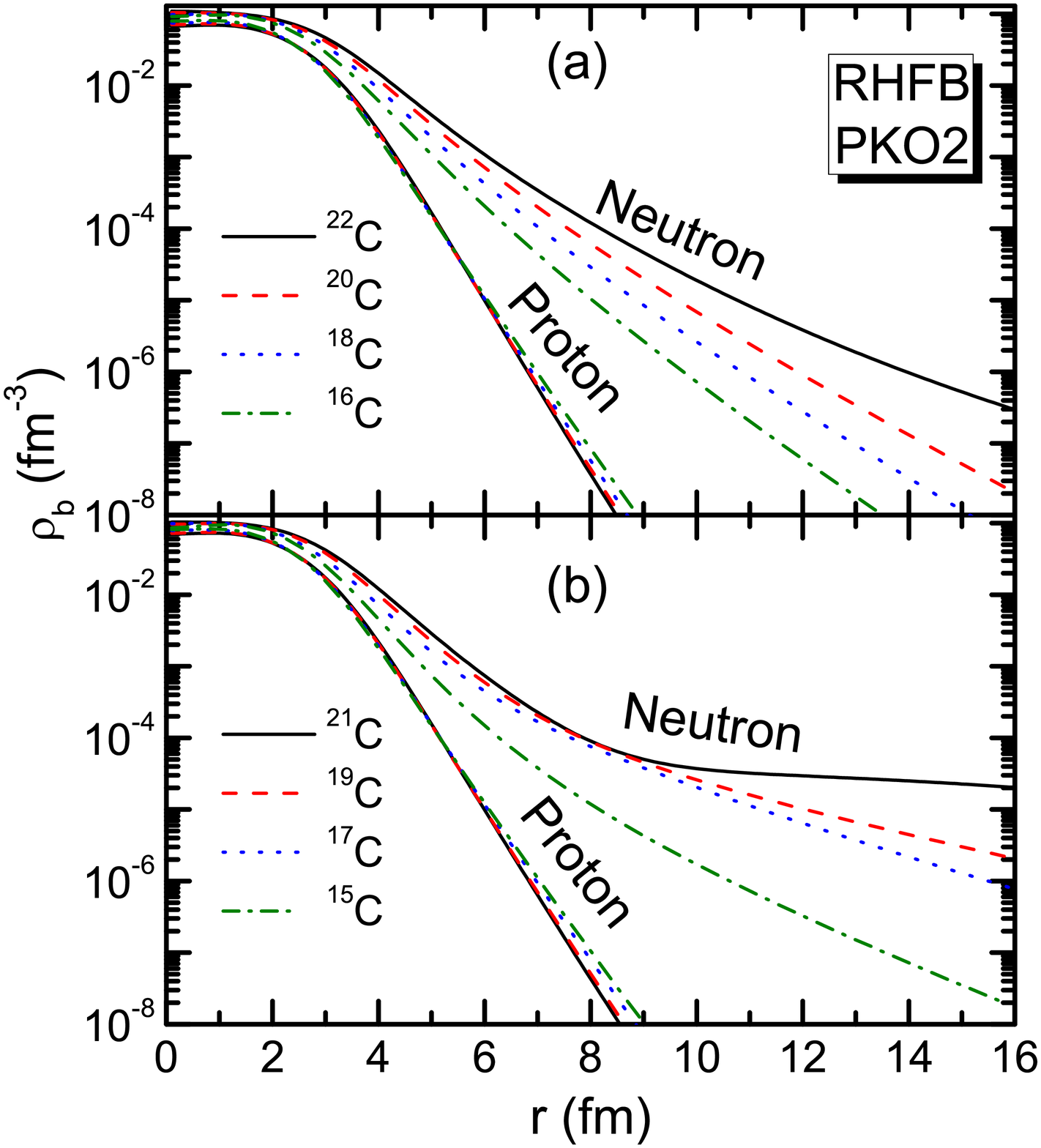}
\fi
\caption{(Color online) Neutron and proton density distributions for even [plot (a)] and odd [plot (b)] Carbon isotopes. The results are calculated by RHFB with PKO2 and the optimized pairing strength factor is adopted as $f=1.10$. See the text for details.} \label{fig:DEN}
\end{figure}

\begin{figure}[htbp]\centering
\ifpdf
\includegraphics[width=0.45\textwidth]{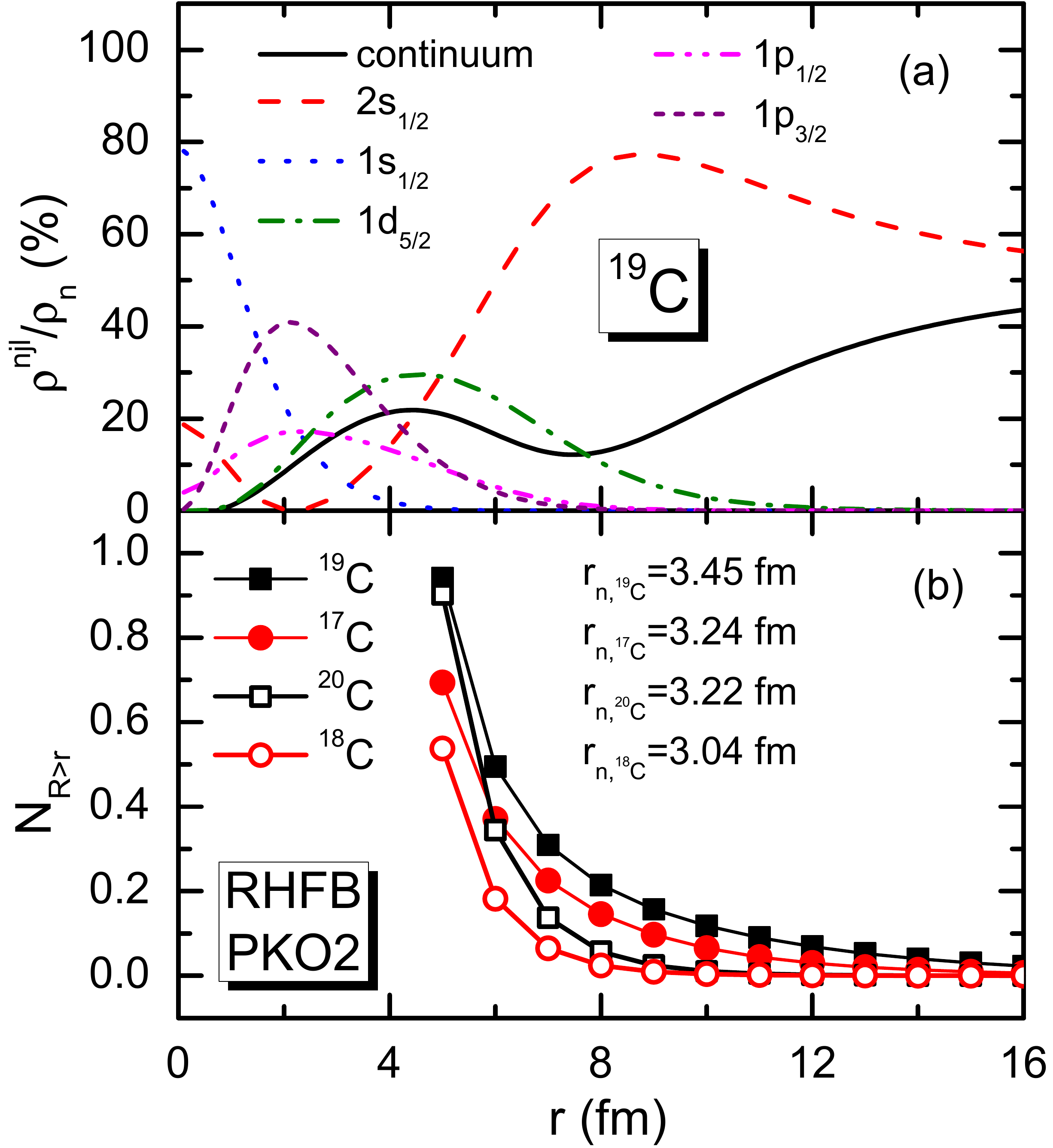}
\else
\includegraphics[width=0.45\textwidth]{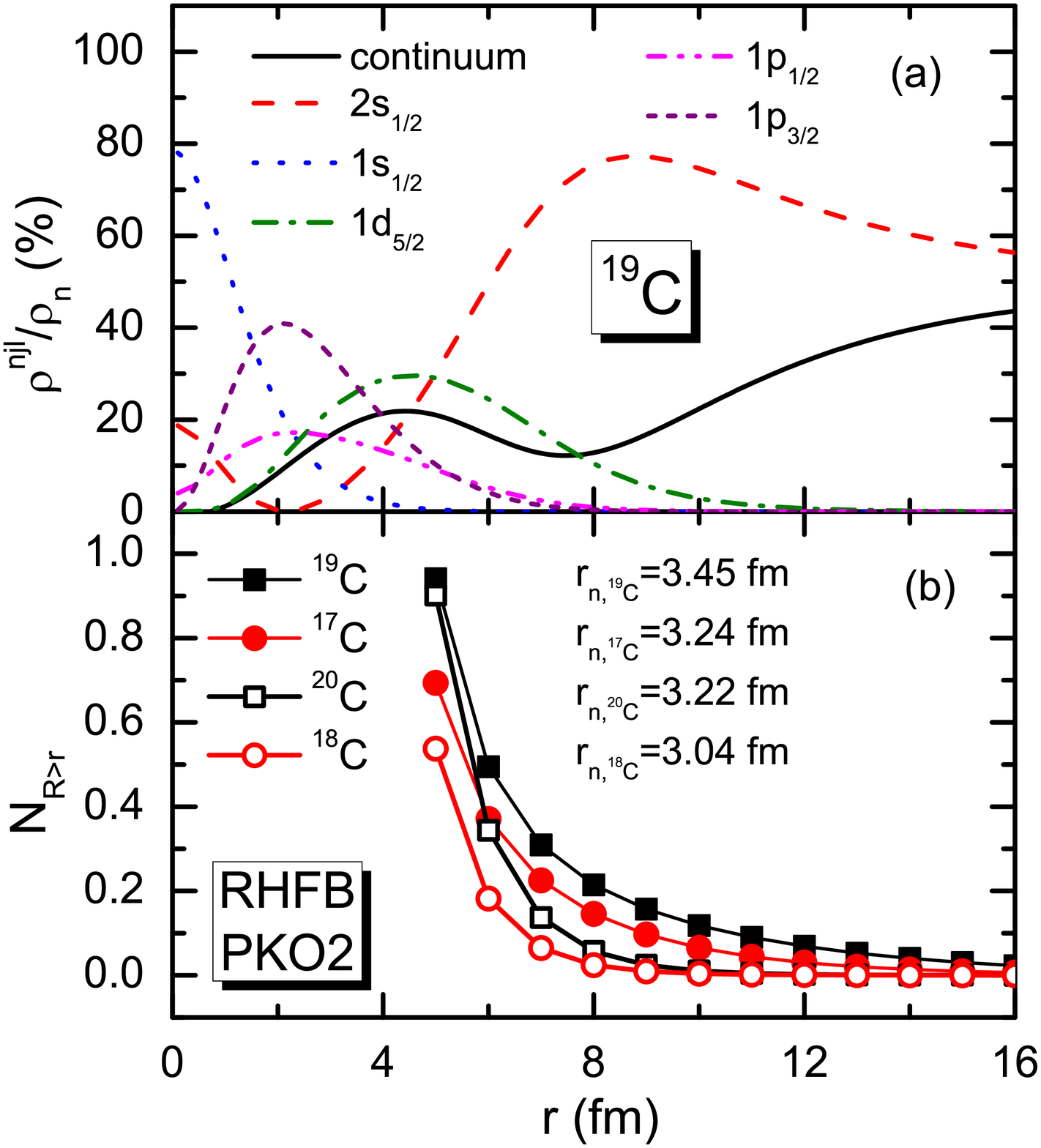}
\fi
\caption{(Color online) (a) Contributions to neutron density ($\rho_n$) from canonical neutron orbits ($\rho^{nlj}$) and the continuum in $^{19}$C, and (b) neutron numbers ($N_{R>r}$) beyond the sphere with radius $r$ for $^{17}$C, $^{18}$C, $^{19}$C and $^{20}$C. The results are extracted from the calculations of RHFB with PKO2. $r_{n, ^{17}\text{C}}$, $r_{n, ^{18}\text{C}}, r_{n, ^{19}\text{C}}$ and $r_{n, ^{20}\text{C}}$ denote the root mean square neutron radii of $^{17}$C, $^{18}$C,$^{19}$C and $^{20}$C, respectively.} \label{fig:VV&halo}
\end{figure}

Aiming at the possible halo structure in Carbon isotopes, Fig. \ref{fig:DEN} shows the neutron and proton density distributions provided by PKO2 calculations for even [plot (a)] and odd [plot (b)] Carbon isotopes. As shown in Fig. \ref{fig:DEN}(a), it seems that the neutron densities of the even isotopes tend to be more and more diffuse, while not distinct enough to support the occurrence of halo structure, when close to the drip line. From the recent data \cite{Audi2011} the two-neutron separation energy of $^{22}$C is 1.56 MeV, which implies that the last two valence neutrons are still bound too deep to spread over a fairly wide range. Hence $^{22}$C may not be a good candidate of well-developed two-neutron halo structure.  Whereas in Fig. \ref{fig:DEN} (b) distinct evidence is presented to demonstrate the halo occurrences in $^{17}$C and $^{19}$C, i.e., more diffused neutron distributions with less neutrons than $^{22}$C. In fact, the strong evidence of the halo occurrence in $^{19}$C can be found from the parallel momentum distribution of $^{18}$C after the breakup of $^{19}$C \cite{Bazin1998}.  As shown in Fig. \ref{fig:S} nearly zero single-neutron separation energies of $^{17}$C and $^{19}$C can be also treated as another evidence for the existence of single-neutron halo structure. For $^{21}$C the negative value of $S_n$ leads to a diverged matter distribution, which might not be a bound nucleus.

\begin{figure}[htbp]\centering
\ifpdf
\includegraphics[width=0.45\textwidth]{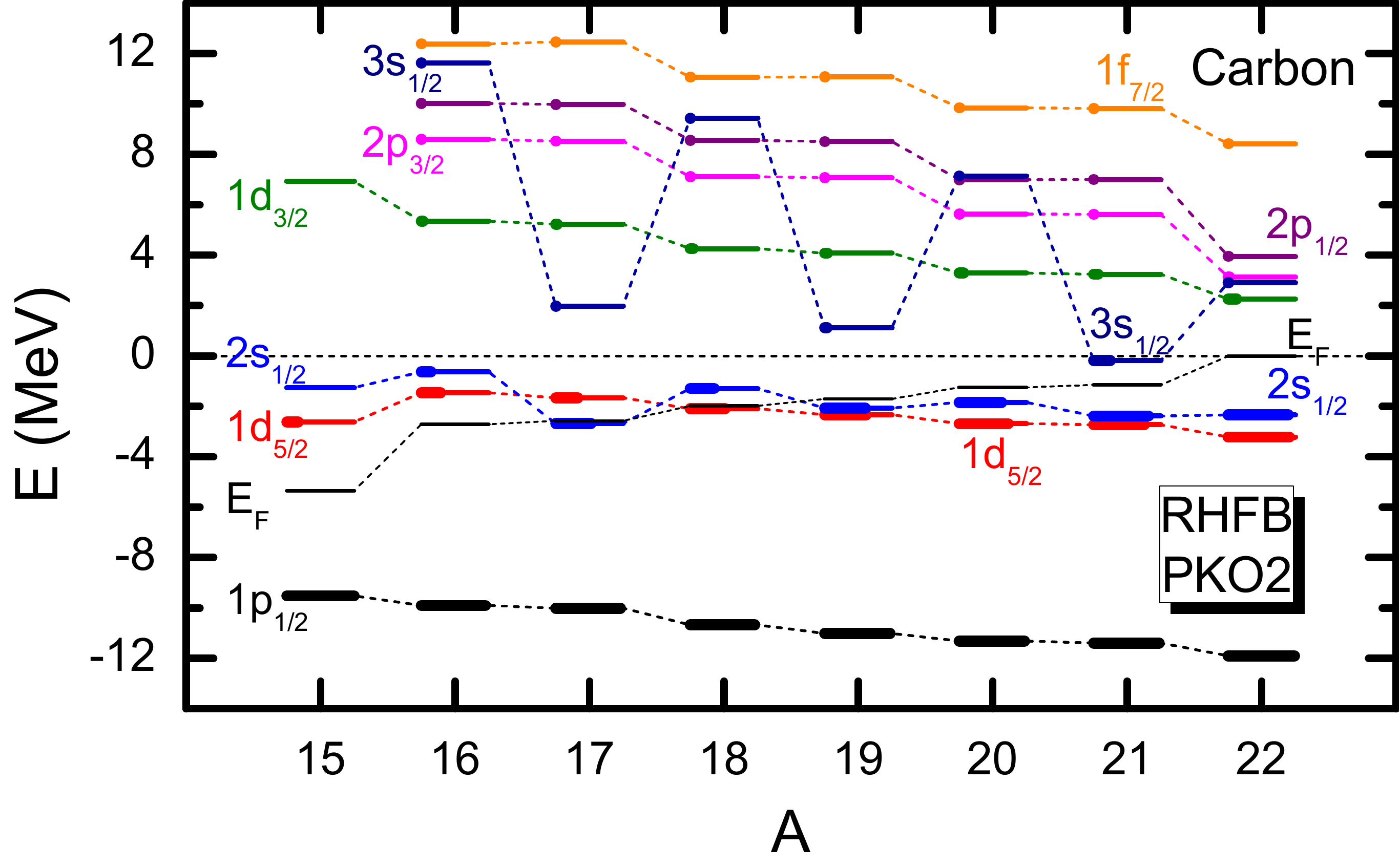}
\else
\includegraphics[width=0.45\textwidth]{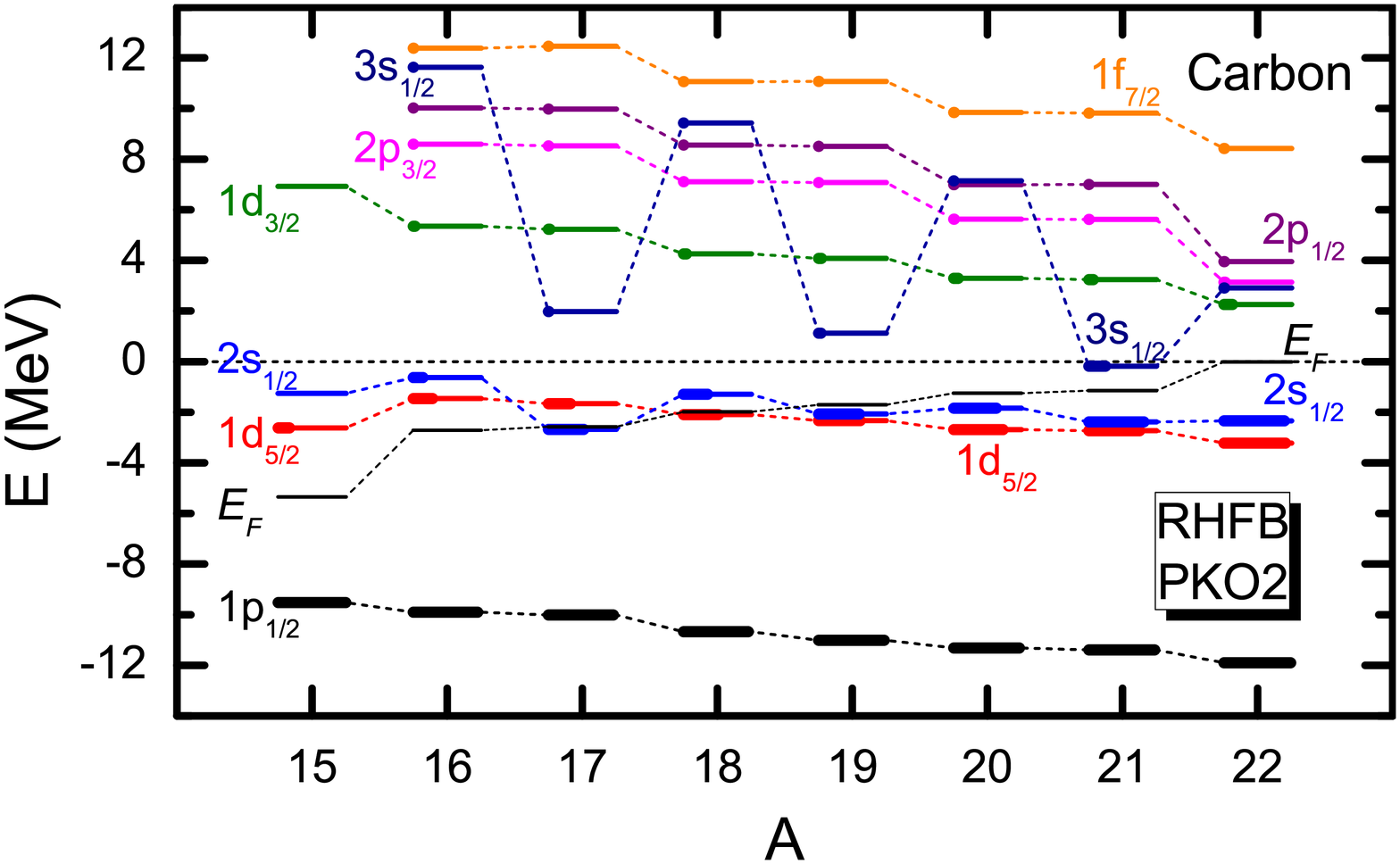}
\fi
\caption{(Color online) Canonical neutron single-particle energies for Carbon isotopes. The results are extracted from the RHFB calculations with PKO2. The length of thick bars corresponds with the occupation probabilities of neutron orbits in half and $E_F$ represent the Fermi levels.} \label{fig:LEV}
\end{figure}

To further illustrate the halo occurrence, Fig. \ref{fig:VV&halo}(a) presents the contributions to the neutron density from different canonical single-particle orbits. It is clearly shown that the dilute matter distribution at large radial distance is dominated by low-$j$ state $2s_{1/2}$ and the continuum, in accordance with the evidence of halo occurrences in $^{11}$Li \cite{Meng1996} and Ca isotopes \cite{Meng2002}. Consistently Fig. \ref{fig:VV&halo}(b) presents another direct evidence, i.e., the number of neutrons $N_{R>r}$ located beyond the sphere with radius $r$. From Fig. \ref{fig:VV&halo}(b) it can be deemed that there exist evident single-neutron halo structures in $^{17,19}$C due to fairly large amount of neutrons spreading far beyond the neutron radii $r_n$. In contrast the values of $N_{R>r}$ in neutron-richer isotopes $^{18,20}$C drop sharply with the increase of radius $r$, consistent with the neutron distributions shown in Fig. \ref{fig:DEN}(a). Combining with the results in Fig. \ref{fig:VV&halo}(a), one can find that both canonical state $2s_{1/2}$ and the continuum present substantial contributions in the formation of halo, while dominated by the formal one due to its zero centrifugal barrier.

As the complemented demonstration, Figure \ref{fig:LEV} shows the neutron canonical single-particle energies for the Carbon isotopes from $^{15}$C to $^{22}$C, where the lengths of the ultra thick bar denote the occupation probabilities in half. From Fig. \ref{fig:LEV} one can find that the valence orbits $2s_{1/2}$ and $1d_{5/2}$ are close to each another and such high level density in general leads to strong pairing effects. Although both valence orbits are fairly close to the continuum limit, the self-consistent RHFB and RHB calculations only support $^{17, 19}$C as the candidates of halo nuclei instead of even drip-line isotope $^{22}$C, which can be well understood from the blocking effects discussed later.

It should be mentioned that the canonical single-particle states in Fig. \ref{fig:LEV} are determined from the diagonalization of the density matrix constructed in the Bogoliubov quasi-particle space. In the calculations of $^{17,19}$C with PKO2, the neutron quasi-particle states near the Fermi surface are blocked, namely the lowest one $1s_{1/2}$. According to the mapping relation between the Bogoliubov quasi-particle and canonical single-particle states \cite{Meng2006, Meng1998NPA}, the corresponding contributions of the blocked quasi-neutron orbits will be mainly mapped into the canonical ones near the Fermi surface, i.e., the canonical $2s_{1/2}$ and $3s_{1/2}$ states as shown in Fig. \ref{fig:LEV}. Compared to the even isotopes, the neutron staying on the canonical orbit $2s_{1/2}$ then becomes much less bound in the odd Carbons due to lacking the extra binding from pairing correlations, which is also illustrated by nearly zero values of $S_n$ in Fig. \ref{fig:S}. As a result, the probability density of the valence state $2s_{1/2}$ tends to be much diffuser than those in even isotopes to develop the halo structure in $^{17,19}$C. Due to the blocking of $s$ orbit, the continuum effects are also enhanced relatively in the odd isotopes because the neutron Cooper pairs in $d_{5/2}$ orbit can be only scattered into the continuum by pairing correlations. In addition, the odd-even staggering on the position of the canonical state $3s_{1/2}$ (see Fig. \ref{fig:LEV}) can be also interpreted by the blocking effects. In the odd Carbon isotopes, the odd quasi-neutron can be mapped partially into the canonical  $3s_{1/2}$ orbits, little while still visible, e.g., $v^2 = 0.034$ for $^{19}$C. As a result relatively enhanced couplings with the core will remarkably lower the $3s_{1/2}$ orbit. Whereas in even isotopes the pairing correlations constrain the valence neutrons spreading mostly over the valence orbits $2s_{1/2}$ and $1d_{5/2}$, and much less neutron can be scattered into the $3s_{1/2}$ states in the continuum, e.g., $v^2 = 0.004$ for $^{20}$C, which therefore become high-lying ones.

As we know, the pairing correlations play significant roles for the halo occurrences in the even nuclear systems, not only in stabilizing nucleus itself but also in developing the halos by scattering the Cooper pairs into the low-lying $s$ or $p$ orbits. The typical examples are $^{11}$Li, the drip line isotopes of Ca, Zr and Ce.
While for the even Carbons particularly $^{22}$C it seems that the extra binding from the pairing correlations makes $s$ orbit too bound to get dilute matter distribution, which also leads to a fairly large two-neutron separation energy $S_{2n}$ (see Fig. \ref{fig:S}). On the contrary for $^{17, 19}$C, due to the absence of extra pairing binding the odd neutron in $s$ orbit presents substantial contribution in the formation of halo structure, which also results in the odd-even staggering (OES) on the neutron radii of Carbon isotopes.

\begin{table}[htbp]
\caption{Matter radii (fm) for Carbon isotopes extracted from the calculations of PKA1, PKO2 PKO3, PKDD, DD-ME2 and PK1, as compared to the experimental data \cite{Liatard1990, Ozawa2001}.} \label{tab:radius}
\begin{tabular}{c|ccccccc}\hline\hline
                       &$^{14}$C&$^{15}$C&$^{16}$C&$^{17}$C&$^{18}$C&$^{19}$C&$^{20}$C \\\hline
PKA1                   &2.53    &  2.74  &  2.73  &  2.89  &  2.91  &  3.02  &  3.08   \\
PKO2                   &2.43    &  2.55  &  2.63  &  2.92  &  2.81  &  3.11  &  2.97   \\
PKO3                   &2.47    &  2.59  &  2.68  &  2.91  &  2.86  &  3.21  &  3.02   \\
PKDD                   &2.43    &  2.75  &  2.66  &  2.91  &  2.86  &  3.26  &  3.03   \\
DD-ME2                 &2.55    &  2.68  &  2.76  &  2.93  &  2.94  &  3.24  &  3.10   \\
PK1                    &2.42    &  2.72  &  2.65  &  2.86  &  2.84  &  3.06  &  3.01   \\ \hline \hline
Ref. \cite{Liatard1990}&2.62\err{(6)} & 2.78\err{(9)}& 2.76\err{(6)}&3.04\err{(11)}&2.90\err{(19)}&2.74\err{(96)}&  $-$    \\ 
\multirow{2}{*}{Ref.
\cite{Ozawa2001}}      &2.30\err{(7)} & 2.48\err{(3)}& 2.70\err{(3)}& 2.72\err{(3)}& 2.82\err{(4)}& 3.13\err{(7)}&2.98\err{(5)}  \\
                       & $-$    & 2.50\err{(8)}&  $-$   & 2.73\err{(4)}&  $-$   & 3.23\err{(8)}& $-$     \\ \hline\hline
\end{tabular}
\end{table}

Before discussing the OES of neutron radii, it is worthwhile to check the quantitative precision for the theoretical description of the radius. In Table \ref{tab:radius} are shown the matter radii of neutron-rich Carbon isotopes obtained from the calculations of PKA1, PKO2, PKO3, PKDD, DD-ME2 and PK1, as compared to the experimental data \cite{Liatard1990, Ozawa2001}. It is found that both RHFB and RHB calculations with the selected effective Lagrangians provide appropriate agreement with the data, which to some extent demonstrates the theoretical reliability.

\begin{figure}[htbp]
\ifpdf
\includegraphics[width=0.45\textwidth]{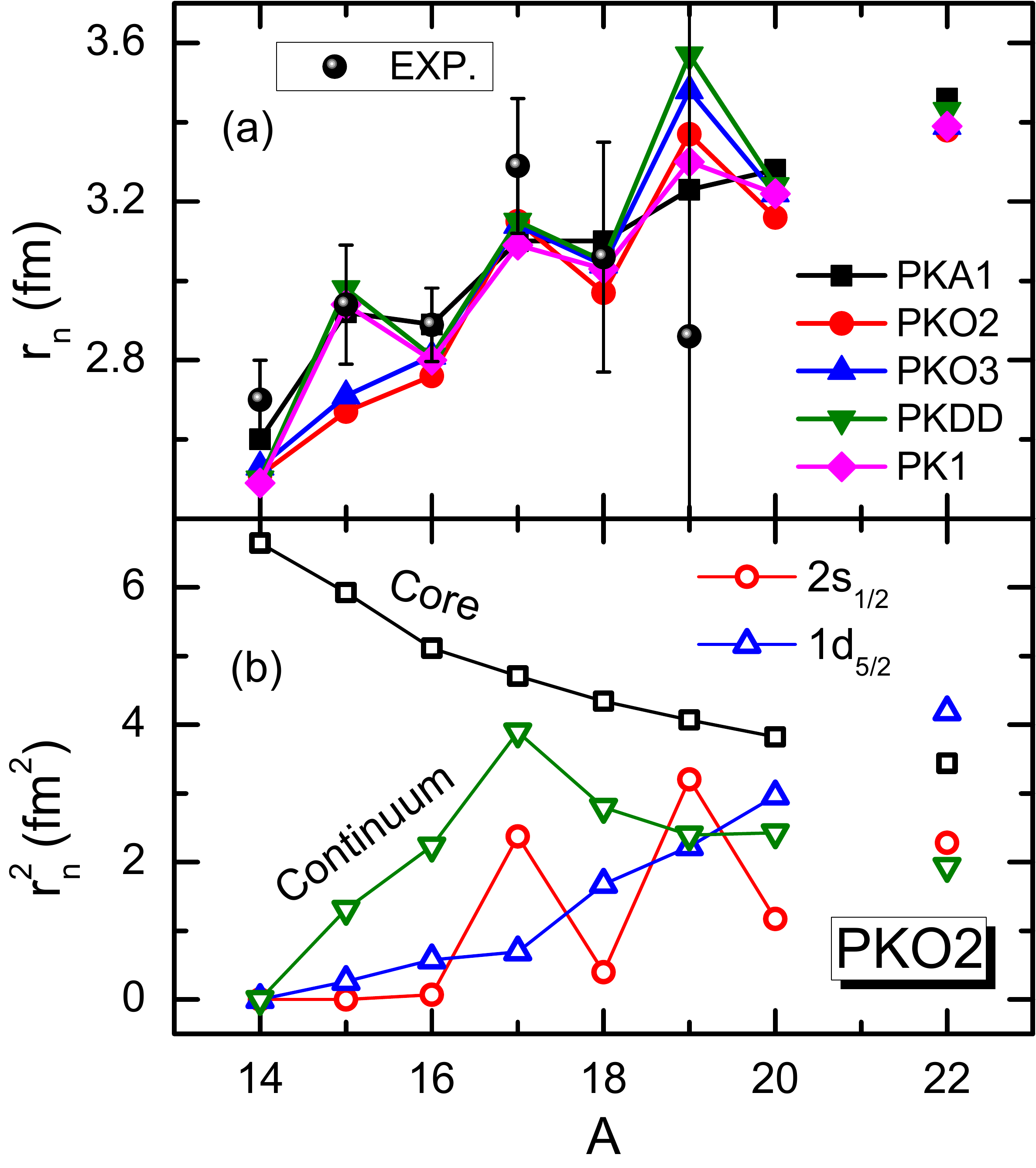}
\else
\includegraphics[width=0.45\textwidth]{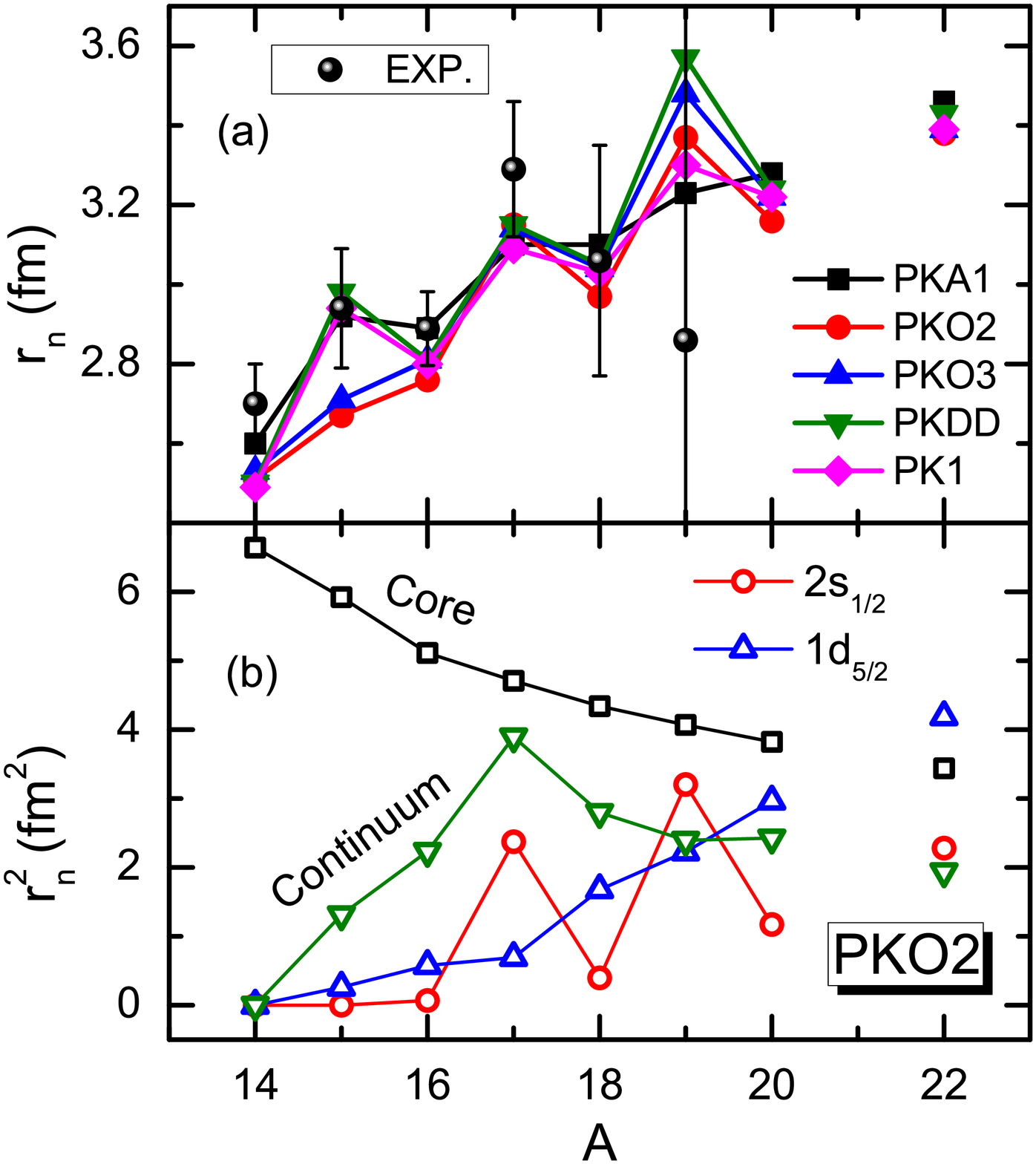}
\fi
\caption{(Color online) (a) Neutron root mean square radii calculated by RHFB with PKA1, PKO2 and PKO3, and by RHB with PKDD and PK1, as compared with the data [for $^{19}$C it reads as 2.86(1.4) fm] \cite{Liatard1990}, and (b) corresponding contributions from the neutron core orbits ($1s_{1/2}$, $1p_{3/2}$ and $1p_{1/2}$), valence orbits ($2s_{1/2}$ and $1d_{5/2}$) and the continuum. The results are provided by the calculations of RHFB with PKO2. See the text for details.}
\label{fig:Rn}
\end{figure}

In fact not only for the total ones, the selected effective Lagrangians with optimized pairing forces also present proper quantitative descriptions for the neutron radii. As shown in Fig. \ref{fig:Rn}(a) and referred to the data \cite{Ozawa2001}, the neutron root mean square radii from $^{14}$C to $^{22}$C are well reproduced by both RHFB and RHB calculations to certain quantitative precision. Evidently as shown in Fig. \ref{fig:Rn}(a) that all the theoretical calculations present distinct OES on the neutron radii, in accordance with the experimental systematics. Specifically, as seen from Fig. \ref{fig:Rn} (b), such OES is determined by the valence neutrons lying in the canonical orbit $2s_{1/2}$ and also depends on the fact whether the corresponding Bogoliubov quasi-particle $s$ orbit is blocked or not. Referring to Table \ref{tab:bl}, one can find that the blocking configurations are consistent with the OES in Fig. \ref{fig:Rn}(a). In $^{15}$C which has larger neutron radius than $^{16}$C in the calculations with PKA1, PKDD and PK1, the quasi-particle $s$ orbit is blocked. Due to similar reason and consistently with the halo occurrence, the neutron radii of halo nuclei $^{17,19}$C are distinctly larger than the even neighbors. The exceptions are the calculations with PKO2 and PKO3 at $^{15}$C and the one with PKA1 at $^{19}$C, where the neutron radii change smoothly. As seen from Table \ref{tab:bl} such exceptional cases correspond with the blocking of $d_{5/2}$ orbit, in which the odd neutron is localized mostly inside the nucleus by the centrifugal barrier. As a result, the ground state of $^{19}$C determined by PKA1 does not correspond with a halo structure since the odd neutron blocks the $d_{5/2}$ orbit and due to the pairing effects the paired neutrons in the low-$j$ $s$ state are bound too strongly to distribute extensively.

\begin{table}[htbp]
\caption{Binding energies in MeV (upper panel) and neutron radii in fm (lower panel) for $^{15, 17, 19}$C calculated by the effective Lagrangians PKA1, PKO2, PKO3, PKDD, DD-ME2 and PK1 with different blocking configurations. For each odd isotopes, the first and second rows correspond with blocking neutron ($\nu$) orbits $s_{1/2}$ and $d_{5/2}$, respectively. The bold types denote the ground states. }   \tabcolsep=0.25em\label{tab:bindings&radii}
\begin{tabular}{crrrrrr}      \hline\hline
                          &   PKA1       &    PKO2     &    PKO3     &      PKDD   &    DD-ME2   &     PK1    \\ \hline
                          & \multicolumn{6}{c}{Binding energies (MeV)}   \\ \hline
\multirow{2}{*}{$^{15}$C} & {\bf$-$107.48} &{   $-$104.98} &{   $-$105.95} &{\bf$-$105.30} &{\bf$-$105.79} &{\bf$-$105.91}\\
                          & {   $-$106.63} &{\bf$-$105.69} &{\bf$-$106.11} &{   $-$105.23} &{   $-$105.35} &{   $-$105.79}\\ \hline
\multirow{2}{*}{$^{17}$C} & {\bf$-$112.99} &{\bf$-$110.32} &{\bf$-$110.81} &{\bf$-$109.83} &{\bf$-$110.11} &{\bf$-$110.76}\\
                          & {   $-$112.44} &{   $-$110.06} &{   $-$110.30} &{   $-$109.30} &{   $-$109.50} &{   $-$110.21}\\ \hline
\multirow{2}{*}{$^{19}$C} & {   $-$117.03} &{\bf$-$114.17} &{\bf$-$114.43} &{\bf$-$113.35} &{\bf$-$113.21} &{\bf$-$114.78}\\
                          & {\bf$-$117.21} &{   $-$113.13} &{   $-$113.65} &{   $-$112.72} &{   $-$112.71} &{   $-$114.16}\\ \hline
                          & \multicolumn{6}{c}{Neutron radii (fm)}   \\ \hline
\multirow{2}{*}{$^{15}$C} & {\bf     2.92} &{        3.05} &{        2.99} &{\bf     2.98} &{\bf     2.93} &{\bf     2.94}\\
                          & {        2.66} &{\bf     2.67} &{\bf     2.71} &{        2.70} &{        2.72} &{        2.88}\\ \hline
\multirow{2}{*}{$^{17}$C} & {\bf     3.10} &{\bf     3.15} &{\bf     3.14} &{\bf     3.15} &{\bf     3.16} &{\bf     3.09}\\
                          & {        3.04} &{        2.91} &{        2.98} &{        2.99} &{        3.04} &{        2.97}\\ \hline
\multirow{2}{*}{$^{19}$C} & {        3.42} &{\bf     3.37} &{\bf     3.48} &{\bf     3.57} &{\bf     4.20} &{\bf     3.31}\\
                          & {\bf     3.23} &{        4.85} &{        3.43} &{        3.31} &{        3.60} &{        3.19}\\\hline\hline

\end{tabular}
\end{table}

As the further illustration of the consistent relation between the OES and blocking configurations, Table \ref{tab:bindings&radii} shows the binding energies and neutron radii of $^{15, 17, 19}$C extracted from the self-consistent calculations with the blockings of $s_{1/2}$ and $d_{5/2}$ orbits, respectively. As shown in the lower panel, the blockings of the low-$j$ $s$ orbit in general lead to more extensive neutron distributions, from which are also well demonstrated the blocking effects in the formation of single-neutron halo structure of $^{17, 19}$C. Specifically for the calculations of $^{19}$C with different blocking, the binding energies determined by PKA1 are close to each another and in fact when $s$ orbit is blocked PKA1 also supports the halo occurrence in $^{19}$C. In contrast the others present distinct differences on the binding energies, especially for PKO2 which confirms the halo emergence in ground state evidentally.

It is well known that pairing correlation plays an important role in stabilizing the finite nuclei, especially the exotic ones. For $^{11}$Li, the neutron drip-line isotopes of Ca, Zr and Ce, it is already demonstrated that the pairing correlations show positive effects in both stabilizing and developing the halo structures. While in the RHFB and RHB calculations of $^{17, 19}$C, the quasi-particle $s$ orbit is blocked and the corresponding contributions are mainly mapped to the canonical orbit $2s_{1/2}$, which plays the dominate role in the single-neutron halo formation of $^{17, 19}$C. This implies that the unpaired odd neutron in low-$j$ orbit may also contribute to develop a halo structure when it is not so deeply bound. From previous analysis it is just due to the lack of extra binding from the pairing correlations that the odd-neutron in $s$ orbit can spread over far beyond the center of nucleus.

\section{Summary and Perspectives}\label{summary}
In this work we have systematically calculated the Carbon isotopes using the relativistic Hartree-Fock-Bogliubov (RHFB) theory with PKA1, PKO2 and PKO3 as well as the relativistic Hartree-Bogliubov (RHB) theory with PKDD, DD-ME2, PK1 and NL2. It is found that with the optimized pairing force the selected effective Lagrangians except NL2 can properly describe the structure properties of the Carbon isotopes, e.g., reproducing the binding energies and matter radii by certain quantitative precision. Specifically the distinct evidences have demonstrated for the single-neutron halo occurrences in $^{17,19}$C, as well as the odd-even staggering (OES) of neutron radii in the vicinity of neutron drip line. While the self-consistent RHFB or RHB calculations do not support the emergence of two-neutron halo structure in $^{22}$C as indicated by the experimental reaction cross section measurement \cite{Tanaka2010}. Further detailed analysis shows that the halo emergences in $^{17, 19}$C, as well as the OES of neutron radii, are essentially concerned with the blocking effects in the odd Carbon isotopes. Different from even nuclear systems, in which the pairing correlations play significant roles in both developing and stabilizing the halo structures, the unpaired odd neutron in weakly bound low-$j$ $s$ orbit dominates the halo formation in $^{17, 19}$C, as well as reproducing the OES of neutron radii for the drip-line Carbon isotopes.


It should be noticed that for the odd Carbons the blocking treatment in this work is just the first-order evaluation of the blocking effects and the current effects induced by the odd neutron are neglected as well. In addition, due to the limit of the present theoretical platform, we only performed the spherical calculations for the Carbon isotopes within the relativistic Hartree and Hartree-Fock theories, while some Carbon isotopes are potentially deformed. After considering the shape fluctuations in both $\beta$ and $\gamma$ deformations, the average neutron quadrupole deformations $\lrb{\langle \beta\rangle_n, \langle \gamma\rangle_n}$ of $^{16,18,20}$C are $(0.50, 21^\circ)$, $(0.49, 29^\circ)$ and $(0.50, 21^\circ)$, respectively \cite{Jiangming2011}. It is then expected that the shape fluctuations will bring some influence on the structure properties of the Carbon isotopes, especially in the vicinity of drip line.  Therefore, the self-consistent treatment of the deformation as well as the odd-particle effects is perspected to be considered carefully for more reliable description of Carbon isotopes.

\begin{acknowledgements}
This work is partly supported by the National Science Foundation of China under Grant Nos. 11075066 and 11205075, the Fundamental Research Funds for the Central Universities under Contracts No. lzujbky-2012-k07 and No. lzujbky-2012-7, and the Program for New Century Excellent Talents in University.
\end{acknowledgements}


\end{document}